\documentclass{article}
\date{October 19, 2008}

\usepackage{amssymb,amsmath}
\usepackage{theorem}
\usepackage{parskip}
\usepackage{fullpage}
\usepackage{hyperref} 

\newtheorem{theorem}{Theorem}
\newtheorem{lemma}[theorem]{Lemma}

\newtheorem{definition}[theorem]{Definition}

\def\squareforqed{\hbox{\rlap{$\sqcap$}$\sqcup$}}
\def\qed{\ifmmode\squareforqed\else{\unskip\nobreak\hfil
\penalty50\hskip1em\null\nobreak\hfil\squareforqed
\parfillskip=0pt\finalhyphendemerits=0\endgraf}\fi}

\newenvironment{proof}{\begin{trivlist}\item[]{\flushleft\bf Proof }}
{\qed\end{trivlist}}

\newcommand{\op}[1]{\mathsf{#1}}

\newcommand{\ket}[1]{\vert #1 \rangle}
\newcommand{\inner}[2]{\langle #1 \vert #2 \rangle}
\newcommand{\proj}{\mathsf{\Pi}}

\newenvironment{keywords}{\begin{trivlist}\item[]{\flushleft\bf Keywords. }}
{\end{trivlist}}

\begin{document}

\title{Exact quantum lower bound for Grover's problem\footnote{This
    work has been supported by Canada's NSERC, the Canadian Institute
    for Advanced Research (CIFAR), and MITACS.}}

\author{%
C\u{a}t\u{a}lin Dohotaru\thanks{%
Department of Computer Science, 
University of Calgary,
2500 University Drive N.W.,
Calgary, Alberta, Canada T2N 1N4.
Email: \texttt{\{dcatalin,hoyer\}@ucalgary.ca}.}
\and 
Peter H{\o}yer\footnotemark[2]
}

\maketitle

\begin{abstract}
  One of the most important quantum algorithms ever discovered is
  Grover's algorithm for searching an unordered set.  We give a new
  lower bound in the query model which proves that Grover's algorithm
  is exactly optimal.  Similar to existing methods for proving lower
  bounds, we bound the amount of information we can gain from a single
  oracle query, but we bound this information in terms of angles.
  This allows our proof to be simple, self-contained, based on only
  elementary mathematics, capturing our intuition, while obtaining at
  the same an exact bound.

  \medskip
\end{abstract}

\section{Introduction}
\label{sec:intro}

Grover's algorithm~\cite{grover1} is one of the most celebrated
quantum algorithms ever devised.  The algorithm and its many
extensions demonstrate that quantum computers can speed up many
search-related problems by a quadratic factor over classical
computers.  The algorithm is based on some of the most fundamental
properties of quantum mechanics and has consequently found uses in a
very wide range of situations, including
unordered searching~\cite{grover1,bbht}, 
communication complexity~\cite{buhrman},
counting~\cite{hoyeramp}, 
cryptography~\cite{ad},
learning theory~\cite{ser}, 
network flows~\cite{ambm},
zero-knowledge~\cite{wat}, and
random walks~\cite{nayak},
just to name a few.  The underlying principles of the algorithm are
very versatile and are readily amendable to a number of variations in
applications.  By clever insight into its basic principles, it has
been adapted to for instance local searching of spatial
structures~\cite{aar}, searching erroneous data~\cite{hoyerberror},
and searching structures with variable search costs~\cite{ambvar}.
Grover's algorithm and its generalizations are in conclusion one of
the most successful frameworks ever discovered for quantum information
processing.

Given these overwhelmingly many positive applications of Grover's
search algorithm, it is natural to ask if Grover's algorithm is
optimal, or if an even faster routine could take its place in the
above applications.  The unanimous answer is that no better algorithm
exists for searching unordered structures on a quantum computer.
Grover's algorithm is in other words optimal.  The optimality of the
algorithm has been established through many different approaches,
including
adversarial arguments~\cite{amb},
degree of polynomials~\cite{beals},
hybrid arguments~\cite{bbbv},
Kolmogorov complexity~\cite{laplante}, and
spectral decompositions~\cite{barnum}.

Common for all the above lower bounds is, however, that they do not
show that Grover's algorithm is exactly optimal, but only
asymptotically optimal.  The above lower bound results on quantum
searching do not exclude the possibility that there might be another
quantum algorithm that solves Grover's search problem say 10\% faster
than Grover's own algorithm.

Fortunately we do know that Grover's algorithm is exactly optimal
through the singular work of Zalka~\cite{zalka}.  By carefully
inspecting each step in Grover's algorithm, Zalka is able to argue
that it is exactly optimal.  Zalka's proof involves defining a certain
function $g$ and proving it has some appropriate behaviour, including
$g'>0$ and $g''<0$, using Lagrange multiplies to solve constrained
extremum problems, and eventually concluding
that four different inequalities are saturated by Grover's algorithm.
Zalka's construction seems to require an intimate understanding
of Grover's algorithm and fluency in finding extrema for various
multi-variate functions.  

Grover and Radhakrishnan~\cite{GR} make several simplifications to
Zalka's proof and construct a more explicit and rigorous proof that
allows them to give a near-tight lower bound for Grover's problem.
Their near-optimal theorem applies without modifications to success
probabilities of at least~0.9, whereas Zalka's original proof applies
to any choice of success probability and any size of search space.

The aim of the present paper is to give a new tight lower bound for
the unordered search problem that is as simple and transparent as
possible, using only elementary mathematics, and that does not presume
any knowledge of Grover's algorithm or any other upper bound.  Our
proof applies, as Zalka's, to any choice of success probability and
any size of search space.

\section{Proving lower bounds for quantum algorithms}
\label{sec:general}
In the unordered search problem, we are given a bitstring $x$ as
input.  The input is given to us as an oracle so that the only
knowledge we can gain about the input is in asking queries to the
oracle.  We are interested in solving the unordered search problem
with the least number of queries to the oracle.  We model the oracle
\begin{equation*}
\op{O}_x \ket{i;\,w} = 
\begin{cases} (-1)^{x_i} \ket{i;\,w} & \textup{ if $1 \leq i \leq N$} \\
   \hphantom{(-1)^{x_i}} \ket{i;\,w} & \textup{ if $i = 0$.}
\end{cases}
\end{equation*}

\begin{definition}[Unordered search problem]
  We are given a bitstring $x \in \{0,1\}^N$ as an oracle, and
  promised that there exists a unique index $1 \leq i \leq N$ for
  which $x_i=1$.  We want to output an index $1 \leq j \leq N$ such
  that $i=j$ with probability at least~$p$.
\end{definition}

In the following, we identify the $N$ possible inputs with the set $y
\in \{1, 2, \ldots, N\}$ so that for instance input $y=7$ denotes the
input bitstring $x$ in which bit~7 is~1 ($x_7=1$) and the $N-1$ other
bits are~0 ($x_i = 0 \textnormal{ for $i \neq 7$}$).

The most straightforward classical \textit{deterministic} algorithm
for solving the unordered search problem would be to simply query the
oracle for the $N-1$ bits $x_1, x_2, \ldots, x_{N-1}$.  If any of
these $N-1$ bits equals~1, we output the corresponding index, and
otherwise, we output the unqueried index $N$.  This algorithm always
outputs the correct answer, uses $N-1$ queries, and is optimal.
An optimal \textit{probabilistic} algorithm is to pick a set of $T=
\lceil p N -1\rceil$ distinct indices uniformly at random, and query
the oracle on those $T$ indices.  If any of the $T$ bits equals~$1$,
we output the corresponding index, and otherwise, we output one of the
remaining $N-T$ indices, picked again uniformly at random.  The
probability we output the correct index is $\frac{T+1}{N} \geq p$.
Grover's algorithm is the best known \textit{quantum} algorithm for
the unordered search problem.  It uses of order $\sqrt{p N}$ queries
and outputs the correct index with probability at least~$p$.

Any quantum algorithm in the oracle model starts in a state that is
independent of the oracle~$x$.  For convenience, we take the start
state to be~$\ket{0}$ in which all qubits are initialized to~0.  It
then evolves by applying arbitrary unitary operators $\op{U}$ to the
system, alternated by queries $\op{O}_x$ to the oracle~$x$, followed
by a conclusive measurement of the final state, the outcome of which
is the result of the computation.  We assume (without loss of
generality) that the final measurement is a von Neumann measurement
represented by a finite set of orthogonal projectors $\big\{ \proj_y
\big\}$ that sum to the identity.  In symbols, a quantum
algorithm~$\op{A}$ that uses $T$ queries to the oracle, computes the
final state
\begin{equation*}
\ket{\Psi_x^T} = \op{U}_T \op{O}_x \op{U}_{T-1} \cdots
  \op{U}_1 \op{O}_x \op{U}_0 \ket{0}
\end{equation*}
which is then measured, yielding the answer $y$ with probability
$\big\| \proj_y \ket{\Psi_x^T} \big\|^2$.  

A~more detailed and excellent introduction to the query model is given
in~\cite{bw}, and a discussion of lower bounds for the model
in~\cite{hoyerlb}.

\section{Exact lower bound for quantum searching}
\label{sec:thebound}

In the unordered search problem, we are given one of $N$ possible
inputs~$x$, and we produce one of $N$ possible outputs~$y \in \{1, 2,
\ldots, N\}$.  For the algorithm to succeed with probability at
least~$p$, we require that $\big\| \proj_x \ket{\Psi_x^T} \big\|^2
\geq p$ for all $x \in \{1, 2, \ldots, N\}$.  Let $\Psi^t = \op{U}_t
\op{O}_u \op{U}_{t-1} \cdots \op{U}_1 \op{O}_u \op{U}_0 \ket{0}$
denote the state after $t$ queries when the oracle $u = 00 \cdots 0$
is the all-zero bitstring, in which case, the oracle $\op{O}_u$ acts
as the identity.  Let $\Psi^{i,T}_y = \op{U}_T \op{O}_u \cdots
\op{O}_u \op{U}_{i+1} \op{O}_u \op{U}_i \op{O}_y \cdots \op{U}_1
\op{O}_y \op{U}_0 \ket{0}$ denote the final state after $T$ queries,
where we use oracle~$y$ for the first $i$ oracle queries and the
identity for the latter $T-i$ oracle queries.

We now give our new exactly optimal lower bound for the unordered
search problem.  We present our proof in parallel with the standard
(asymptotically optimal) hybrid argument lower bound derived
from~\cite{bbbv} which seems to be the simplest of the existing lower
bounds.  Both proofs require three steps, and we present each of these
steps in a form so that the two proofs resemble each other as closely
as possible and are as simple as possible.

The key to our new lower bound is the use of angles as opposed to
distances as in the standard proof in~\cite{bbbv}.  We define the
\emph{quantum angle} between two non-zero vectors as
\begin{equation}
  \measuredangle (\psi, \psi') = \arccos 
  \frac{\big\vert \inner{\psi}{\psi'}\big\vert}{\| \psi \| \| \psi' \|}.
\end{equation}
This seems to be the most appropriate definition of angles for quantum
computing and can be readily generalized to mixed states through
fidelity and satisfies, in particular, the triangle inequality.

\subsection{First step}
The first step in the proof is to establish a Cauchy-Schwarz-like
inequality (for each of our two measures, distances and angles) which
will allow us to bound the amount of information we can learn by each
individual query.

\begin{lemma}[Cauchy--Schwarz --- distance version]\label{lm:cs}
\begin{equation}
\max\left\{ \sum\limits_{i=1}^{N} a_{i} \mid 0 \leq a_{i}
  \textup{ and }
  \sum\limits_{i=1}^{N} a_{i}^{2} \leq 1 \right\} = \sqrt{N}.
\end{equation}
\end{lemma}

\begin{proof}
  First note that when all $a_{i}$'s are equal, the maximum value of
  the sum is~$\sqrt{N}$.  Now, assume that $\sqrt{N}$ is not the
  maximum value of the sum.  Then there exist $N$ numbers
  $b_{1},\ldots,b_{N}$ for which the maximum is attained.  At least
  two of the $b_{i}$'s are not equal, denote them by $x$ and~$y$.
  Replacing both $x$ and $y$ with their average, the sum we want to
  maximize remains unchanged, while the sum of squares strictly
  decreases since
\begin{equation*}
x^{2}+y^{2}-2\left( \frac{x+y}{2}\right) ^{2}=\frac{1}{2}(x-y)^{2}>0.
\end{equation*}
We can thus increase all $b_i$'s by a tiny amount while keeping the
sum of squares at most~1, contradicting the assumption that the
$b_i$'s attain the maximum.  It follows the maximum is attained when
all $a_i$'s are equal.
\end{proof}

\begin{lemma}[Cauchy--Schwarz --- angle version]\label{lm:cs:angle}
\begin{equation}
\max\left\{ \sum\limits_{i=1}^{N} \theta_{i} 
  \mid 0 \leq \theta_{i} \leq \frac{\pi}{2}
  \textup{ and }
  \sum\limits_{i=1}^{N} \sin^2 \theta_{i} \leq 1 \right\} 
  = N \arcsin \frac{1}{\sqrt{N}}.
\end{equation}
\end{lemma}

\begin{proof}
  First note that when all $a_{i}$'s are equal, the maximum value of
  the sum is~$N \arcsin \frac{1}{\sqrt{N}}$.  Now, assume that this is
  not the maximum value of the sum.  Then there exist $N$ angles
  $\varphi_{1},\ldots,\varphi_{N}$ for which the maximum is attained.
  At least two of the $\varphi_{i}$'s are not equal, denote them by
  $u$ and~$v$.  Replacing both $u$ and $v$ with their average, the sum
  we want to maximize remains unchanged, while the sum of squares
  strictly decreases since\footnote{The equality can be proven by
    showing that both sides are equal to $\cos(u+v) -
    \frac{1}{2}\cos(2u) - \frac{1}{2}\cos(2v)$, or by applying Euler's
    formula.}
\begin{equation*}
  \sin ^{2}u+\sin ^{2}v-2\sin ^{2}\left( \frac{u+v}{2}\right) =2\sin
  ^{2}\left( \frac{u-v}{2}\right) \cos (u+v)>0.
\end{equation*}
We can thus increase all $\varphi_i$'s by a tiny amount while keeping
the sum of squares at most~1, contradicting the assumption that the
$\varphi_i$'s attain the maximum.  It follows the maximum is attained
when all $\theta_i$'s are equal.
\end{proof}

\subsection{Second step}
The second step is then to show that the amount of information we
learn by each of the $T$ query can only add up linearly (with respect
to our two measures, distances and angles).

\begin{lemma}[Increase in distance by $T$ queries] 
\label{lm:avgchange}
  The average distance after $T$ queries is at
  most~$2T\frac{1}{\sqrt{N}}$.
\end{lemma}

\begin{proof}
  We have, using the triangle inequality,%
\begin{eqnarray*}
  \frac{1}{N}\sum\limits_{y=1}^{N}
  \big\Vert \Psi^{T} - \Psi^{T}_{y}\big\Vert 
  &=& \frac{1}{N}\sum\limits_{y=1}^{N}
  \big\Vert \Psi^{T,T}_{y} - \Psi^{0,T}_{y}\big\Vert 
  \;\leq\;  \frac{1}{N}\sum\limits_{y=1}^{N} \sum\limits_{i=0}^{T-1}
  \big\Vert \Psi^{i+1,T}_{y} - \Psi^{i,T}_{y}\big\Vert \\
  &=&\frac{1}{N}\sum\limits_{i=0}^{T-1} \sum\limits_{y=1}^{N}
  \big\Vert \op{O}_{y}\Psi^{i} - \Psi^{i}\big\Vert 
  \;=\;  \frac{1}{N}\sum\limits_{i=0}^{T-1} \sum\limits_{y=1}^{N}
  2\big\Vert \proj_y \Psi^{i} \big\Vert \\
  &\leq& 2\sum\limits_{i=0}^{T-1} \frac{1}{\sqrt{N}} 
  \;=\; 2T\frac{1}{\sqrt{N}},
\end{eqnarray*}%
where the last inequality follows from the inequality proven in
Lemma~\ref{lm:cs}.
\end{proof}

\begin{lemma}[Increase in angle by $T$ queries] 
\label{lm:avgchange:angle}
  The average angle after $T$ queries is at most $2T \Theta$, where
  $\Theta = \arcsin(\frac{1}{\sqrt N})$.
\end{lemma}

\begin{proof}
  We have, using the triangle inequality for angles,
\begin{eqnarray*}
  \frac{1}{N}\sum\limits_{y=1}^{N}
  \measuredangle \big( \Psi^{T}, \Psi^{T}_{y} \big) 
  &=& \frac{1}{N}\sum\limits_{y=1}^{N}
  \measuredangle \big( \Psi^{T,T}_{y}, \Psi^{0,T}_{y} \big)
  \;\leq\;  \frac{1}{N}\sum\limits_{y=1}^{N} \sum\limits_{i=0}^{T-1}
  \measuredangle \big( \Psi^{i+1,T}_{y}, \Psi^{i,T}_{y}\big) \\
  &=& \frac{1}{N}\sum\limits_{i=0}^{T-1} \sum\limits_{y=1}^{N}
  \measuredangle \big( \op{O}_{y}\Psi^{i}, \Psi^{i}\big)
  \;=\;  \frac{1}{N}\sum\limits_{i=0}^{T-1} \sum\limits_{y=1}^{N}
  \arccos \big( \big\vert \cos (2\theta_{y}^{i}) \big\vert \big) \\
  &\leq&   \frac{1}{N}\sum\limits_{i=0}^{T-1}\sum\limits_{y=1}^{N} 2\theta_{y}^{i}
  \;\leq\; 2\sum\limits_{i=0}^{T-1} \Theta
  \;=\;    2T\Theta,
\end{eqnarray*}
where the last inequality follows from the inequality for angles
proven in Lemma~\ref{lm:cs:angle}.
\end{proof}

\subsection{Third step}
The third and final step is then to show that by the end of the
algorithm, after all $T$ queries, our measure (distance or angle,
respectively) is large.

\begin{lemma}[Distinguishability of final states --- distance version]
\label{lm:final}
Suppose that the algorithm correctly outputs $y$ with probability at
least~$p$ after $T$ queries, given oracle $\op{O}_y$.  Then the
average distance is at least
\begin{equation}
  \frac{1}{N}\sum\limits_{y=1}^{N}
  \big\Vert \Psi^{T} - \Psi^{T}_{y}\big\Vert 
  \geq \frac{1}{\sqrt 2} \Big( 1 + \sqrt{p} - \sqrt{1-p} - \frac{2}{\sqrt N}\Big).
\end{equation}
\end{lemma}

\begin{proof}
  The distance after $T$ queries is at least
\begin{eqnarray*}
  \big\Vert \Psi^{T} - \Psi^{T}_{y} \big\Vert 
  &\geq& \frac{1}{\sqrt 2} \Big(
  \big\Vert \proj_y \Psi^T_{y} - \proj_y \Psi^T \big\Vert
  + \big\Vert \proj_y^{\perp} \Psi^T - \proj_y^{\perp} \Psi^T_{y} \big\Vert \Big) \\
  &\geq& \frac{1}{\sqrt 2} \Big(
    \big\Vert \proj_y \Psi^T_{y} \big\Vert 
  - \big\Vert \proj_y \Psi^T \big\Vert 
  + \big\Vert \proj_y^{\perp} \Psi^T \big\Vert 
  - \big\Vert \proj_y^{\perp} \Psi^T_{y} \big\Vert  \Big) \\
  &\geq& \frac{1}{\sqrt 2} \Big(
    \sqrt{p}
  - \big\Vert \proj_y \Psi^T \big\Vert 
  + \big\Vert \proj_y^{\perp} \Psi^T \big\Vert  
  - \sqrt{1-p} \Big)\\
  &\geq& \frac{1}{\sqrt 2} \Big(
    \sqrt{p}
  - \sqrt{1-p}
  + 1
  - 2 \big\Vert \proj_y \Psi^T \big\Vert  \Big),
\end{eqnarray*}
where the first inequality follows from the inequality $(a-b)^2 \geq
0$, the second-last inequality from the success probability being at
least~$p$, and the other two from the triangle inequality.  The
average distance after $T$ queries is thus at least
\begin{eqnarray*}
  \frac{1}{N}\sum\limits_{y=1}^{N} 
    \big\Vert \Psi^{T} - \Psi^{T}_{y} \big\Vert
  &\geq& \frac{1}{N}\sum\limits_{y=1}^{N} 
  \frac{1}{\sqrt 2} \Big( 1 + \sqrt{p} 
      - \sqrt{1-p} - 2 \big\Vert \proj_y \Psi^T \big\Vert \Big) \\
  &=&
  \frac{1}{\sqrt 2} \Big( 1 + \sqrt{p} - \sqrt{1-p} 
  - \frac{2}{N} \sum\limits_{y=1}^{N} 
    \big\Vert \proj_y \Psi^T \big\Vert \Big) \\
  &\geq&
  \frac{1}{\sqrt 2} \Big( 1 + \sqrt{p} - \sqrt{1-p} - \frac{2}{\sqrt N}\Big),
\end{eqnarray*}
where the last inequality follows from the inequality proven in
Lemma~\ref{lm:cs}.
\end{proof}

\begin{lemma}
[Distinguishability of final states --- angle version]
\label{lm:final:angle}
Suppose that the algorithm correctly outputs $y$ with probability at
least~$p$ after $T$ queries, given oracle $\op{O}_y$.  Then the
average angle is at least
\begin{equation}
  \frac{1}{N}\sum\limits_{y=1}^{N}
  \measuredangle \big( \Psi^{T},\Psi^{T}_{y}\big)
  \geq \Theta^{T}-\Theta,
\end{equation}
where $\sin^2(\Theta^T) = p$ and $\sin^2(\Theta) = \frac{1}{N}$.
\end{lemma}

\begin{proof}
  The angle difference after $T$ queries is at least
\begin{eqnarray*}
  \measuredangle \big(\Psi^{T}, \Psi^{T}_{y}\big) 
  &=& \arccos \big( \big\vert \langle \Psi^{T} \vert \Psi^{T}_{y}\rangle \big\vert\big) \\
  &=& \arccos \big( \big\vert \langle \Psi^{T} \vert (\proj_y + \proj_y^{\perp}) \vert
  \Psi^{T}_{y}\rangle \big\vert\big) \\
  &\geq& \arccos \big( 
  \big\Vert \proj_y \Psi^{T} \big\Vert \cdot
  \big\Vert \proj_y \Psi^{T}_{y} \big\Vert +
  \big\Vert \proj_y^{\perp} \Psi^{T} \big\Vert \cdot
  \big\Vert \proj_y^{\perp} \Psi^{T}_{y} \big\Vert \big)\\
  &=&\arccos \big( \sin \theta_{y}^{T}\sin \phi_{y}^{T} +
  \cos \theta_{y}^{T} \cos \phi_{y}^{T}\big) \\
  &=&\arccos \big(\cos (\phi_{y}^{T} - \theta_{y}^{T})\big) \\
  &=&\big\vert \phi_{y}^{T} - \theta_{y}^{T}\big\vert,
\end{eqnarray*}%
where $\sin(\phi_y^T) = \Vert\proj_y \Psi_y^T\Vert$ and
$\sin(\theta_y^T) = \Vert\proj_y \Psi^T\Vert$.  The average angle
difference after $T$ queries is thus at least
\begin{equation*}
  \frac{1}{N}\sum\limits_{y=1}^{N} \measuredangle \big( \Psi^{T}, \Psi^{T}_{y}\big) 
  \;\geq\; \frac{1}{N}\sum\limits_{y=1}^{N} \big( \phi_{y}^{T} - \theta_{y}^{T}\big) 
  \;\geq\; \Theta^{T}-\frac{1}{N} \sum\limits_{y=1}^{N} \theta_{y}^{T} 
  \;\geq\; \Theta^{T}-\Theta, 
\end{equation*}
where the second-last inequality follows from the success probability
being at least~$p$, and the last inequality from the inequality for
angles proven in Lemma~\ref{lm:cs:angle}.
\end{proof}

\subsection{Concluding the proof}
 
Since each of our two measures is 0 initially, is large by the end of
the algorithm, and can only increase modestly by each query, we can
conclude that a large number of queries is required.

\begin{theorem}[Asymptotic lower bound for searching --- distance
  version] \label{thm:cs} 

  The unordered search pro\-blem with success probability $p>0$ requires
  at least $T \geq \frac{\sqrt{N}}{2\sqrt 2} \Big( 1 + \sqrt{p} -
  \sqrt{1-p} - \frac{2}{\sqrt N}\Big)$ queries.
\end{theorem}

\begin{theorem}[Tight lower bound for searching --- angle
  version] \label{thm:cs:angle} 

  The unordered search problem with success probability $p =
  \sin^2(\Theta^T)>0$ requires at least $T \geq \frac{\Theta^T -
    \Theta}{2 \Theta}$ queries.
\end{theorem}

In the case of distances, we conclude that Grover's algorithm is
asymptotically optimal, and in the case of angles, that Grover's
algorithm is exactly optimal.  No other algorithm can achieve even a
constant additive improvement with respect to the number of queries
required for a given success probability.  Compared to other lower
bounds, and even to the hybrid argument, our proof seems surprisingly
simple.  It would be interesting to extend our method to obtain both
simpler and better lower bounds for other problems, and also to find
other uses of quantum angles.


\begin{thebibliography}{99}
%
\bibitem{aar} S. Aaronson and A. Ambainis. 
\emph{Quantum search of spatial regions}.
Theory of Computing, 1(1):47--79, 2005.

%
\bibitem{ad} M. Adcock, R. Cleve, K. Iwama, R. Putra, and S. Yamashita. 
\emph{Quantum lower bounds for the Goldreich-Levin problem}. 
Information Processing Letters, 97(5):208--211, 2006.

%
\bibitem{amb} A. Ambainis. 
\emph{Quantum lower bounds by quantum arguments}. 
Journal of Computer and System Sciences, 64:750--767, 2002.

%
\bibitem{ambvar} A. Ambainis. 
\emph{Quantum search with variable times}.
Proceedings of the 25th Annual 
Symposium on Theoretical Aspects of Computer Science,
Lecture Notes in Computer Science 49--61, 2008.

%
\bibitem{ambm} A. Ambainis and R. \v{S}palek. 
\emph{Quantum algorithms for matching and network flows}. 
Proceedings of the 23rd Annual 
Symposium on Theoretical Aspects of Computer Science,
Lecture Notes in Computer Science 3884:172--183, 2006.

%
\bibitem{barnum} H. Barnum, M. Saks, and M. Szegedy. 
\emph{Quantum decision trees and semidefinite programming}. 
Proceedings of the 18th IEEE Conference on Computational Complexity, 
pp. 179--193, 2003.

%
\bibitem{beals} R. Beals, H. Buhrman, R. Cleve, M. Mosca, and R. de~Wolf. 
\emph{Quantum lower bounds by polynomials}. 
Journal of the ACM, 48(4):778--797, 2001.

%
\bibitem{bbbv} H. Bennett, E. Bernstein, G. Brassard, and U. Vazirani. 
\emph{Strengths and weeknesses of quantum computing}. 
SIAM Journal on Computing, 26(5):1510--1523, 1997.

%
\bibitem{bbht} M. Boyer, G. Brassard, P. H\o yer, and A. Tapp. 
\emph{Tight bounds on quantum searching}. 
Fortschritte Der Physik, 46(4--5):493--505, 1998.

%
\bibitem{hoyeramp} G. Brassard, P. H\o yer, M. Mosca, and A. Tapp. 
\emph{Quantum amplitude amplification and estimation}. 
In Quantum Computation and Quantum Information: 
{A}~Millennium Volume, AMS Contemporary Mathematics Series, 
Volume 305, 2002.

%
\bibitem{buhrman} H. Buhrman, R. Cleve, and A. Wigderson. 
\emph{Quantum vs.{} classical communication and computation}. 
Proceedings of the 30th ACM Symposium on Theory of Computing, pp 63--65, 1998. 

%
\bibitem{bw} H. Buhrman and R. de~Wolf.
\emph{Complexity Measures and Decision Tree Complexity: {A}~Survey}.
Theoretical Computer Science, 288(1):21--43, 2002. 

%
\bibitem{grover1} L. K. Grover. 
\emph{Quantum mechanics helps in searching for a needle in a haystack}. 
Physical Review Letters, 79(2):325--328, 1997.

%
\bibitem{GR} L. K. Grover and J. Radhakrishnan. 
\emph{Is partial quantum search of a database any easier?}
Proceedings of the 17th Annual ACM Symposium on Parallelism 
in Algorithms and Architectures,
pp 186--194, 2005.

%
\bibitem{hoyerberror} P. H\o yer, M. Mosca, and R. de Wolf. 
\emph{Quantum search on bounded-error inputs}. 
Proceedings of the 30th International
Colloquium on Automata, Languages, and Programming, 
Lecture Notes in Computer Science 2719:291--299, 2003.

%
\bibitem{hoyerlb} P. H\o yer and R. \v{S}palek. 
\emph{Lower bounds on quantum query complexity}. 
Bulletin of the European Association for Theoretical Computer Science, 
87:78--103, 2005.

%
\bibitem{laplante} S. Laplante and F. Magniez. 
\emph{Lower bounds for randomized
and quantum query complexity using Kolmogorov arguments}.
SIAM Journal on Computing, 38(1):46--62, 2008.

%
\bibitem{nayak} F. Magniez, A. Nayak, J. Roland, and M. Santha. 
\emph{Search via quantum walk}. 
Proceedings of the 39th ACM Symposium on Theory of Computing, pp 575--584, 2007.

%
\bibitem{ser} R. Servedio and S. Gortler. \emph{Equivalences and separations
between quantum and classical learnability}. 
SIAM Journal on Computing, 33(5):1067--1092, 2004.

%
\bibitem{wat} J. Watrous. \emph{Zero-knowledge against quantum attacks}.
To appear in SIAM Journal on Computing, 2008.

%
\bibitem{zalka} Ch. Zalka. 
\emph{Grover's quantum searching is exactly optimal}. 
Physical Review A, 60:2746--2751, 1999.


\end{thebibliography}
\end{document}